\documentclass[aps,prl,twocolumn,superscriptaddress, showpacs]{revtex4-1}

\usepackage{graphicx}
\usepackage{bm}
\usepackage{amssymb}
\usepackage{pifont}

\begin{document}


\title{Weak measurement of the Dirac distribution}



\author{Jeff S. Lundeen}

\email{jeff.lundeen@nrc-cnrc.gc.ca}

\affiliation{Institute for National Measurement Standards, National Research Council,
1200 Montreal Rd., Ottawa, ON, K1A 0R6}

\author{Charles Bamber}

\affiliation{Institute for National Measurement Standards, National Research Council,
1200 Montreal Rd., Ottawa, ON, K1A 0R6}

\begin{abstract}
Recent work {[}J.S. Lundeen et al. Nature, \textbf{474}, 188 (2011){]}
directly measured the wavefunction by weakly measuring a variable
followed by a normal (i.e. {}`strong') measurement of the complementary
variable. We generalize this method to mixed states by considering
the weak measurement of the product of the two observables, which,
as a non-Hermitian operator, is normally unobservable. This generalized
method provides mixed states an operational definition related to
the operator representation proposed by Dirac. Uniquely, it can be
performed {}`in situ', determing the quantum state without destroying
it.
\end{abstract}
\pacs{03.65.Ta, 03.65.Wj, 42.50.Dv, 03.67.-a}


\maketitle

The wavefunction $\Psi$ is at the heart of quantum mechanics, yet
its nature has been debated since its inception. It is typically relegated
to being a calculational device for predicting measurement outcomes.
Recently, Lundeen et al. proposed \cite{Lundeen2011, Lundeen2006} a simple and
general operational definition of the wavefunction based on a method
for its direct measurement: {}``it is the average result of a weak
measurement of a variable followed by a strong measurement of the
complementary variable.'' The {}`wavefunction' referred to here
was introduced along with the Schr�dinger Equation and can be thought
of as a special case of a quantum state, known as a {}`pure state.'
The general case is represented by the density operator $\rho$, which
can describe both pure and {}`mixed' states. The latter incorporates
both the effects of classical randomness (e.g., noise) and entanglement
with other systems (e.g., decoherence). The density operator plays
an important role quantum statistics, quantum information, and the
study of decoherence. As quantified by the Purity $\mu=\mathrm{Tr}[\rho^{2}]\leq1$,
a density operator can range from pure ($\mu=1$) to completely mixed
($\mu=1/N$, where $N$ is the Hilbert space dimension). We investigate
what becomes of our operational definition of the wavefunction when
it is applied to general quantum states. 

The standard method for experimentally determining the density operator
is Quantum State Tomography \cite{Vogel1989,*Smithey1993,*Breitenbach1997,*White1999}.
In it, one makes a diverse set of measurements on an ensemble of identical
systems and then determines the quantum state that is most compatible
with the measurement results. An alternative is our direct measurement
method, which may have advantages over tomography, such as simplicity,
versatility, and directness.  As compared to tomography, which works with mixed states, a significant limitation of the direct
measurement method is that it has only
been shown to work with pure states. The most common goal of state
determination is to evaluate a system's difference from a target quantum
state, which might be, for example, a potential resource for a quantum
information protocol. Since the target is usually a pure state, this
difference often due to mixedness, and so it can be quantified by
the purity of the state.

In this paper, we consider whether our direct method can be generalized
to density operators. Previous works have developed direct methods
to measure quasi-probability distributions such as the Wigner function
\cite{Wigner1932}, Husimi Q-function \cite{Husimi1940}, and the
Glauber-Sudarshan P-function \cite{Glauber1963,*Sudarshan1963}. These
are position-momentum (i.e. {}`phase-space') distributions that
are equivalent to the density operator, and have many, but not all,
of the properties of a standard probability distribution. The most
used quasi-probability distribution is the Wigner function $W_{\rho}(x,p)$,
which has the following properties: (1) it is real; (2) less than
one, $\left|W_{\rho}\right|\leq1$; (3) its marginals give correct
predictions for position $x$ and momentum $p$ probability distributions
(i.e. $\int W_{\rho}(x,p)dx=\mathrm{Prob}(p)$); and (4) the mean
value of an observable $A$ is just an overlap, $\left\langle A\right\rangle =2\pi\hbar\int\int W_{\rho}(x,p)W_{A}(x,p)dxdp$.
However, unlike a standard probability $W_{\rho}$ can be negative.
This is what enables it to be compatible with the predictions of quantum
mechanics. The Wigner function can be directly measured \cite{Banaszek1996}
by displacing the system in phase space and then measuring the parity
operator (this is a nontrivial requirement, see \cite{Laiho2010}).
Equivalently, the integral of the interference between a pair of rotated
and displaced replicas of the system will give the Wigner function
\cite{Mukamel2003,*Smith2005}. The Husimi Q-function can be directly
measured by an eight-port homodyne apparatus or by projection on the
harmonic oscillator ground state \cite{Leonhardt1997,*Kanem2005}.
These phase-space distributions are created to be the closest quantum
analogs to a classical probability distribution. In this sense, they
are inherently amenable to direct measurement.

A lesser known phase-space distribution that can be used to represent
a quantum operator was introduced by Dirac in 1945 in his paper, {}``On
the Analogy Between Classical and Quantum Mechanics.'' The Dirac
distribution has since been extended to discrete Hilbert spaces \cite{Chaturvedi2006}.
In various guises it has been investigated periodically during last half-century \cite{Johansen2007}. In optics, variations of the Dirac distribution have been used widely,
appearing in Walther's definition of the radiance function in radiometry
\cite{Walther1968} and Wolf's specific intensity \cite{Wolf1976}
(as pointed out in Ref. \cite{Chaturvedi2006}). Dirac showed that
an operator $A$ could be represented by its overlap with the basis
states of two non-commuting variables. Specifically, one can represent
an operator in phase-space as $S_{A}(x,p)=\left\langle x\right|A\left|p\right\rangle \cdot\left\langle p|x\right\rangle $.
If $A=\rho$, the Dirac distribution is representation of the quantum
state of a system and shares features (2) to (4), from above, with
the Wigner function \cite{Chaturvedi2006}. Unlike the Wigner function, it is compatible with Bayes' law and, thus, is consistent with a quantum analog of classical determinism \cite{Hofmann2011a}.  Hence, it has many of
the desired features of quasi-probability distributions but also has a novel logical consistency.

However, the Dirac distribution has two peculiar features: one, it
is sensitive to the ordering of the non-commuting variables (i.e.
$\left\langle x\right|A\left|p\right\rangle \cdot\left\langle p|x\right\rangle \neq\left\langle p\right|A\left|x\right\rangle \cdot\left\langle x|p\right\rangle $);
and two, Dirac noted that although his distribution {}``was developed
to provide a formal probability'' for $x$ and $p$ {}``it turns
out to be in general a complex number.'' Surprisingly, Dirac saw
the sensitivity of his distribution to the arbitrary choice of ordering
not as a drawback but, rather, as a desirable characteristic, since
it emphasized the key difference between quantum and classical mechanics:
non-commutivity of operators \cite{Moyal2006}. Even accepting this,
the Dirac distribution's complex nature has likely inhibited its acceptance
as a quasi-probability distribution. Other complex phase-space operator
representations have proven useful in quantum optics \cite{Drummond1980,*Walls2008}.
And ways in which complex probabilities can be made logically consistent
with a frequentist interpretation of probability have been outlined
\cite{Youssef1994}. Surprisingly, despite being complex we will show
that the Dirac distribution is directly measurable, much like the
Wigner function. Notably, in our method, the choice of variable ordering
has a clear physical meaning. Thus, it removes a degree of arbitrariness
from the distribution's definition.

We begin by considering what happens to our method for directly measuring
the wavefunction when the state is not pure. At the heart the direct
method is weak measurement, which we now review. Upon a standard (i.e.{}`strong')
measurement of a quantum system, the density operator changes depending
on the measurement outcome. This change can be viewed in two ways:
as a straightforward update of our knowledge of the state of the system
or, at least in some situations \cite{Mir2007}, as a physical disturbance
due to the interaction between the system and the measurement apparatus
(e.g. the Heisenberg Microscope \cite{Wheeler1983}). To perform a
weak measurement one reduces the strength of the interaction between
the measured system and the measurement apparatus. This results in
a commensurate reduction in the amount of information one receives
from the measurement and, also, a reduction in the disturbance induced
by the measurement. This tradeoff is inherent in quantum mechanics:
measurement precision for system disturbance \cite{Fuchs1996,*Fuchs2001,*Banaszek2001}.
While a weak measurement on a single system provides little information,
by repeating it on an arbitrarily large ensemble of identical systems
one can precisely determine the \textit{average} measurement result
with arbitrary precision. This average is simply the standard quantum
expectation value, $\left\langle \Psi\right|A\left|\Psi\right\rangle $,
where $A$ is the observable measured and $\left|\Psi\right\rangle $
is the pure state of the system. This is true independent of the strength
of the measurement.

Over the last decade, interest in weak measurement has grown as researchers
have realized its potential for interrogating quantum systems in a
coherent manner \cite{Aharonov2010,*Cho2011}. Weak measurement theory
has been used to model and understand photonic phenomena in birefrigent
photonic crystals \cite{Solli2004}, fiber networks \cite{Brunner2003,*Brunner2004},
cavity QED \cite{Wiseman2002}, and quantum tunelling \cite{Steinberg1995,*Steinberg1995a}.
Weak measurement provides insight into a number of fundamental quantum
effects, including the role of the uncertainty principle in the double-slit
experiment \cite{Wiseman2003,*Mir2007}, the Legget-Garg inequality
\cite{Williams2008,*Palacios-Laloy2010,*Goggin2011}, the quantum box
problem \cite{Resch2004}, and Hardy's paradox \cite{Lundeen2009}.
Weak measurement has also been used to amplify small experimental
effects \cite{Hosten2008,*Dixon2009,*Feizpour2011} and as feedback
for control of a quantum system \cite{Smith2004,*Gillett2010}. Weak
measurements have been demonstrated in both classical \cite{Ritchie1991}
and non-classical systems \cite{Pryde2005}.

A distinguishing feature of weak measurement is that, in the limit
of zero interaction, the quantum state of the system remains unchanged.
Subsequent measurements can now provide additional information about
that initial quantum state. Consider a subsequent strong measurement
of observable $C$ that results in outcome $c$ (corresponding to
eigenstate eigenstate $\left|c\right\rangle $). The average result
of the weak measurement of $A$ in the sub-ensemble of systems giving
$C=c$ is called the {}`Weak Value' and is given by \cite{Aharonov1988},
\begin{equation}
\left\langle A\right\rangle _{\Psi}^{c}=\frac{\left\langle c\right|A\left|\Psi\right\rangle }{\left\langle c|\Psi\right\rangle }.\label{eq:weakvalue}
\end{equation}
Surprisingly, the weak value can be outside the range of the eigenvalues
of $A$ and can even be complex \cite{Aharonov1990,Lundeen2005,Jozsa2007}.
The concept of weak measurement is universal \cite{Oreshkov2005, Hofmann2010, Dressel2010}
but insight into the meaning of complex weak values can be gained
by considering a particular model of measurement by Von Neumann \cite{Von1955}.
In this general model, the measurement apparatus has a pointer that
shifts in position to indicate the result of a strong measurement
of $A$. In weak measurement, not only does the average position of
pointer shift but also the average momentum of the pointer. These
shifts are proportional to the real and imaginary parts of the weak
value, respectively, thus giving them straightforward physical manifestations
\cite{Aharonov1990,Lundeen2005,Jozsa2007}. The complex nature of
the weak value is what enables us to directly measure real and imaginary
parts of the wavefunction and, as we show later, directly measure
the Dirac distribution.

We now review our method for the direct measurement of the wavefunction.
The concept is general, however here we consider the case of a discrete
Hilbert space. In this space, one is free to choose the basis $\left\{ \left\vert a\right\rangle \right\} $
(associated with observable $A$) in which the wavefunction will be
measured. The method consists of weakly measuring a projector in this
basis $\pi_{a}\equiv\left\vert a\right\rangle \left\langle a\right\vert $,
and post-selecting on a particular value $b_{0}$ of the complementary
observable $B$. By {}`complementary' we mean that $\left\langle a|b_{0}\right\rangle =1/\sqrt{N}$
for all $a$, where $N$ is the dimension of the Hilbert space. That
is, the overlap is real and constant as function of $a$. The existence
of state $\left|b_{0}\right\rangle $ is guaranteed by the existence
of at least two mutually unbiased bases (MUB) in any Hilbert space
\cite{Durt2010}. Using Eq. (\ref{eq:weakvalue}), the quantum state
$\left|\Psi\right\rangle $ is given by
\begin{equation}
\left\vert \Psi\right\rangle =v\cdot\sum\limits _{a}\left\langle \pi_{a}\right\rangle _{\Psi}^{b_{0}}\left\vert a\right\rangle ,
\end{equation}
where $\left\langle \pi_{a}\right\rangle _{\Psi}^{b_{0}}$ is the
weak value and $v$ is a constant that is independent of $a$. Thus
by stepping through the values of $a$ in a series of weak measurements
one can directly measure $\left\vert \Psi\right\rangle $ represented
in the $a$ basis. 

To extend our method to mixed states we need to calculate the weak
value of a system described by a density operator. There always exists
an orthonormal basis $\left\{ \left|\lambda\right\rangle \right\} $
in which,
\begin{equation}
\rho=\sum_{\lambda}p_{\lambda}\left|\lambda\right\rangle \left\langle \lambda\right|.\label{eq:density}
\end{equation}
Thus, a mixed state can always be thought of the random preparation
of a set of orthogonal pure states $\left|\lambda\right\rangle $,
each with probability $p_{\lambda}.$ Refs. \cite{Romito2010,*Haapasalo2011,*Wu2011}
showed that the weak value will simply be a weighted sum of the weak
values for every pure state in the decomposition given in Eq. (\ref{eq:density}):
\begin{equation}
\left\langle A\right\rangle _{\rho}^{c}=\sum_{\lambda}P\left(\lambda|c\right)\left\langle A\right\rangle _{\lambda}^{c},
\end{equation}
where $P\left(\lambda|c\right)$ is the probability of the initial
state being prepared in $\left|\lambda\right\rangle $ given the system
was post-selected in $\left|c\right\rangle $. Using Bayes' Theorem
to simplify this (as in \cite{Romito2010,*Haapasalo2011,*Wu2011}) we
arrive at:
\begin{equation}
\left\langle A\right\rangle _{\rho}^{c}=\frac{\left\langle c\right|A\rho\left|c\right\rangle }{\left\langle c\right|\rho\left|c\right\rangle }.\label{eq:wv_mixed}
\end{equation}
And applying this to our direct measurement method we find,
\begin{equation}
\left\langle \pi_{a}\right\rangle _{\rho}^{b_{0}}=\frac{\left\langle b_{0}|a\right\rangle \left\langle a\left|\rho\right|b_{0}\right\rangle }{\left\langle b_{0}\right|\rho\left|b_{0}\right\rangle }.\label{eq:direct_density}
\end{equation}
Examination of Eq. (\ref{eq:direct_density}) shows that only $2N$
real parameters are found by scanning $a$. This will not generally
be sufficient to determine all the parameters in $\rho$, which has
$N^{2}-1$ real parameters. As might be suspected, our method for
the direct measurement of the wavefunction cannot be used to determine
a mixed state.

We now consider whether the direct measurement technique can be used
for a more modest task: to determine whether a state is mixed. This
could be used as an indicator for when the method should not be used.
Or, if it is also a measure of state purity $\mu$, it could be used
to determine the quality of prepared quantum states. The weak value
in Eq. (\ref{eq:direct_density}) is equal to sum of one row of the
density matrix, $\rho_{a_{1}a_{2}}=\left\langle a_{1}\left|\rho\right|a_{2}\right\rangle $,
i.e. $\left\langle \pi_{a_{1}}\right\rangle _{\rho}^{b_{0}}=\frac{1}{N\cdot\rho_{b_{0}b_{0}}}\sum_{a_{2}}^{N}\rho_{a_{1}a_{2}}$,
where $\rho_{b_{0}b_{0}}$is the probability that the measurement
of $B=b_{0}$. In effect, one measures a distinct weak value for each
pure state in the statistical decomposition in Eq. (\ref{eq:density}).
This simply adds noise to our determination of the weak value for
the density operator. In the limit of zero interaction strength, the
weak measurement exhibits maximal statistical uncertainty and so additional
noise would be imperceptible. Thus, we conclude that the original
direct measurement method does not contain any signature of the purity
of the state. 

We now consider what happens if one replaces the strong measurement
of $B$ with a weak measurement. Specifically, we investigate the
weak measurement of the product of projectors from the two MUB, $S_{ab}\equiv\bigl|b\bigr\rangle\bigl\langle b|a\bigr\rangle\bigl\langle a\bigr|$.
Since there is no longer any post-selection, this average is calculated
in the same manner as the standard quantum expectation value:
\begin{eqnarray}
\left\langle S_{ab}\right\rangle _{\rho} & = & \mathrm{Tr}\left[S_{ab}\rho\right]\label{eq:jointweak-1}\\
 & = & \left\langle a\right|\rho\left|b\right\rangle \left\langle b|a\right\rangle \\
 & = & S_{\rho}(a,b),
\end{eqnarray}
where $S_{\rho}(a,b)$ is the discrete Hilbert space version of the
Dirac distribution as defined in \cite{Chaturvedi2006} (there exist
analogous discrete Wigner functions \cite{Leonhardt1995}). This is
the main result of the paper; A joint weak measurement of complementary
variables $A$ and $B$ directly measures the Dirac distribution at
values $a$ and $b$. By scanning $a$ and $b$, so as to measure
the Dirac distribution for over all values of $(a,b)$, one completely
determines the density operator. Similarly, a joint weak measurement
of $S_{xp}\equiv\bigl|p\bigr\rangle\bigl\langle p|x\bigr\rangle\bigl\langle x\bigr|$
on a mixed state $\rho$ gives the phase-space version of the Dirac
distribution, $S_{\rho}(a,b)$.

In order to calculate the density operator from the Dirac distribution
one must know $\left\langle b|a\right\rangle =\exp\left(i\theta_{ab}\right)/\sqrt{N}$.
Unfortunately, other than in a select few cases, it is not generally
known what are the bases in the MUB set (for any given Hilbert space).
Thus, a general formula for $\theta_{ab}$ is also unknown. However,
if $\left\{ \left\vert a\right\rangle \right\} $ is taken to be the
standard basis (i.e. $\sum_{a=0}^{N}\left|a\right\rangle \left\langle a\right|=I,$
the identity operator) then one MUB, which we take to be $\left\{ \left\vert b\right\rangle \right\} $,
will always be the Fourier basis \cite{Bengtsson2007,Durt2010}, $\left\vert b\right\rangle =\sum_{a=0}^{N-1}\left\vert a\right\rangle \exp\left(i2\pi ab/N\right)/\sqrt{N}$.
In this case, $\theta_{ab}=-2\pi ab/N$, where $a$ and $b$ are integers
solely used to enumerate the states such that $0\leq a,b\leq N-1$.
With these choices for our complementary bases the density operator
is simply related to the Dirac distribution by a Discrete Fourier
Transform,
\begin{equation}
\rho_{a_{1}a_{2}}=\sum_{b=0}^{N-1}S_{\rho}\left(a_{1},b\right)e{}^{i2\pi b(a_{1}-a_{2})/N}.\label{eq:density_transform}
\end{equation}
This explicitly shows that average result of the joint weak measurement,
$\left\langle S_{ab}\right\rangle _{\rho}$, contains the same information
as the density operator. 

However, there is no need to ever transform to the density operator
as all of the system's properties can be found directly with the Dirac
distribution, $S_{\rho}(a,b)$. It can be used to directly calculate
the average result of measuring an observable $O$ on state $\rho$:
\begin{equation}
\left\langle O\right\rangle =N\cdot\sum_{a,b=0}^{N-1}S_{\rho}\left(a,b\right)S_{O}\left(a,b\right)^{*},
\end{equation}
where $S_{O}\left(a,b\right)$ can be calculated by Eq. (\ref{eq:jointweak-1}),
and {*} is the complex conjugate. With $O$ chosen to be the identity
operator $I$, we find that $\sum_{a,b=0}^{N-1}S_{\rho}\left(a,b\right)=Tr\left[\rho\right]=1$.
In other words, the Dirac distribution is normalized in the same manner
as a probability distribution. With $O=\rho$, we find the purity
$\mu=N\cdot\sum_{a,b=0}^{N-1}\left|S_{\rho}\left(a,b\right)\right|^{2},$
which reaffirms that purity is a global property of the density operator
and thus, we are unable to measure purity without completely determing
$\rho$. And finally, with $O=\left\vert a\right\rangle \left\langle a\right\vert $
or $\left\vert b\right\rangle \left\langle b\right\vert $, we see
that the $a$ and $b$ marginals are equal to the probability distributions
of outcomes $a$ and $b$, e.g. $\sum_{a=0}^{N-1}S_{\rho}\left(a,b\right)=\mathrm{Prob}(b)$.
Consequently, the result of our joint weak measurement, the Dirac
distribution of the density operator, is a capable alternative to
the standard quantum quasi-probability distributions, such as the
Wigner function. Its peculiarity is that it is complex whereas probabilities
are real. Nonetheless, our method for directly measuring it provides
an operational meaning to both its real and imaginary parts; they
appear right on our measurement apparatus, in the shifts in the two
conjugate observables of the pointer, e.g., $x$ and $p$. 

One cause for concern in our method is that $S_{ab}$ is not Hermitian
(i.e. $F^{\dagger}=F$, where $\dagger$ indicates the Hermitian Adjoint).
According to the postulates of quantum mechanics this means it can
not be an observable \cite{Shankar1994}. The typical justification
for this postulate is that only a Hermitian operator has real eigenvalues.
This is critical for two reasons: One, Hamiltonians containing these
observables must be Hermitian in order to conserve probability. This
argument has been challenged recently in regards to PT symmetry, \cite{Bender2003,*Ruter2010}.
And two, because we only ever record real values in experiments. An
established counter-example to this justification is the weak value,
Eq. (\ref{eq:weakvalue}), which can be complex and yet can be measured
straightforwardly. One way to view the familiar complex values in
quantum mechanics (e.g., in the Schr�dinger equation) is as a mechanism
for mathematical simplification, enabling the combination of coupled
equations into one equation. In the spirit of this view, the complex
weak value is a just a simplified expression for two real average
measurement outcomes. Complex eigenvalues create no inconsistencies
in this simplified formulation. Consequently, instead of Hermitian
we contend that weakly measured observables are only required to be
Normal, $FF^{\dagger}=F^{\dagger}F$.

The above discussion assumes that the weak measurement of the non-Hermitian
operator $S_{ab}$ is actually possible. It is reasonable to question
whether, in principle, there is a measurement apparatus that can jointly
implement a weak measurement of $\left\vert a\right\rangle \left\langle a\right\vert $
and $\left\vert b\right\rangle \left\langle b\right\vert $? The answer
is yes. One possibility is to have the same measurement pointer coupled
to both $A$ and $B$ variables \cite{Molmer2001,*Aharonov2002}. Another
possibility is to weakly measure $a$ and $b$ projectors separately,
coupling each to its own measurement pointer, and then evaluate the
average correlations in the two pointers (e.g., $\left\langle x_{1}p_{2}\right\rangle $
and $\left\langle x_{1}x_{2}\right\rangle $)\cite{Lundeen2005,Resch2004a,*Mitchison2007,*Mitchison2008}.
Both options have been demonstrated experimentally \cite{Lundeen2009,Yokota2009}.
Either possibility allows one to find the average result of the weak
measurement of $S_{ab}$.

We now return to the operator ordering in the Dirac distribution,
which as it stands is chosen arbitrarily. An examination of Eq. (\ref{eq:jointweak-1})
reveals that in our definition of $S_{ab}$ the measurement of $A$
precedes the measurement of $B$. One might hypothesize that the order
in which two weak measurements conducted would not matter since the
wavefunction is nominally unchanged by both measurements. In Ref.
\cite{Mitchison2008}, the authors considered sequential weak measurements
of observables $E$ and $F$ and three possibilities were distinguished:
measure $E$ then $F$; measure $F$ then $E$; and, measure $E$
and $F$ simultaneously. By analyzing each of these with the Von Neumann
model of measurement, they found that each had a different weak value
if $E$ and $F$ do not commute. In the context of our direct measurement
of the Dirac distribution, the operator ordering choice is, thus,
set by the choice of time-ordering in which one conducts the measurements
of $A$ and $B$. 

Experimentally implementing a joint weak measurement can be challenging
since the signal to noise of the weak measurement of $n$ operators
$\mathrm{SNR}(n)$ goes as $\mathrm{SNR}(1)^{n}.$ We now outline
an easier method to determine the density matrix. It consists of scanning
the post-selected state in our original method for the direct measurement
of the wavefunction. In this scenario, we replace $b_{0}$ with $b$
in Eq. (\ref{eq:direct_density}), where $\left\{ \left\vert b\right\rangle \right\} $
is the Fourier basis. The weak value is given by, $P\left(b\right)\left\langle \pi_{a}\right\rangle _{\rho}^{b}=S_{\rho}(a,b)$,
where the probability to post-select $B=b$ is $P\left(b\right)=\rho_{bb}$.
To consider the experimental meaning of the left-hand side of this
equation, we return to the general formula for the weak value, Eq.
(\ref{eq:wv_mixed}). In the von Neuman model, the pointer shift,
which indicates the weak value of a general observable $A$, is determined
by measuring the average of some pointer observable $D=\sum_{i}d_{i}\left\vert d_{i}\right\rangle \left\langle d_{i}\right\vert $
(e.g. $D=X$) in the subset of systems with outcome $C=c$ or $\left\langle A\right\rangle _{\rho}^{c}=k\sum_{i}d_{i}P\left(d_{i}|c\right)$,
where $k$ is a proportionality constant accounting for the strength
of the measurement. Then using Bayes' Theorem, we find $\left\langle c\right|A\rho\left|c\right\rangle =k\sum_{i}d_{i}P\left(d_{i},c\right)$,
where $P\left(h,g\right)$ means the probability of getting both $h$
and $g$ outcomes in one trial. Applying this result to the weak measurement
of $\left\vert a\right\rangle \left\langle a\right\vert $ while post-selecting
$B=b$ we find,
\begin{equation}
S_{\rho}(a,b)=k\sum_{i}d_{i}P\left(d_{i},c\right).
\end{equation}
Consequently, again we can directly measure the Dirac distribution,
though not through the usual evaluation of the weak value. Instead
one records the joint probability for outcome $c$ on the system of
interest and outcome $d_{i}$ on the pointer, when weak measuring
$\left\vert a\right\rangle \left\langle a\right\vert $. In this way
we can completely determine the density operator through Eq. (\ref{eq:density_transform}).
This determination has a key advantage over standard quantum state
tomography in that it only requires measurements in two of the system's
bases.

In summary, by measuring a pair of complementary variables of a system
it is possible not only to directly measure its wavefunction but,
also, to determine its density operator. To measure over the extent
of wavefunction we only need to scan the first variable. To measure
the density operator we must, additionally, scan the second variable.
Whereas the first variable must be measured weakly, the second can
either be strongly or weakly measured. The latter option corresponds
to the measurement of the product of projectors for two complementary
observables, such as $x$ and $p$. This product operator is a non-Hermitian
operator and thus has no corresponding observable relevant for standard
measurement. Nonetheless, it is weakly measurable, the average result
of which is a two-dimensional complex distribution that was first
introduced by Dirac. With the distribution one can directly predict
the outcome of any measurement on the system through a simple overlap
integral, much the same as how we use the Wigner function for predictions.
An advantage that this method has over other state determination techniques
is that the amount of state disturbance can be minimized. Thus, in
principle, we can characterize quantum states in situ, for instance,
in the middle of quantum computation circuits, or during chemical
reactions, without disturbing the process in which they feature.
\begin{acknowledgments}
We thank Aephraim Steinberg for useful discussions.
\end{acknowledgments}
\bibliography{gen_wavefunction_prl}

\end{document}